\documentclass{aa}
\usepackage{graphicx}
\usepackage{psfig}

\def\etal{{et\,al. }}

\def\etal{et\,al.}

\def\degs{\ifmmode ^{\circ}\else$^{\circ}$\fi}
\def\amin{\ifmmode ^{\prime}\else$^{\prime}$\fi}
\def\asec{\ifmmode ^{\prime\prime}\else$^{\prime\prime}$\fi}
\def\farcs{\hbox{$.\!\!^{\prime\prime}$}}  
\def\fm{\hbox{$.\!\!^{\rm m}$}}            
\def\fd{\hbox{$.\!\!^{\rm d}$}}            
\newbox\grsign \setbox\grsign=\hbox{$>$}
\newdimen\grdimen \grdimen=\ht\grsign
\newbox\laxbox \newbox\gaxbox
\setbox\gaxbox=\hbox{\raise.5ex\hbox{$>$}\llap
     {\lower.5ex\hbox{$\sim$}}}\ht1=\grdimen\dp1=0pt
\setbox\laxbox=\hbox{\raise.5ex\hbox{$<$}\llap
     {\lower.5ex\hbox{$\sim$}}}\ht2=\grdimen\dp2=0pt

\def\m{$^{\rm m}$}
\def\rxj{RX\,J0704.2+6203}
\begin{document}

   \thesaurus{08     
              (08.02.1;  
               08.02.4;  
               08.14.2;  
               08.09.2;  
               13.25.5;  
               02.13.1)} 
   \title{The New AM Her   System RX J0704.2+6203 }

     \subtitle{Northern  Twin of BL Hyi}

   \author{G.\,H.~Tovmassian
        \inst{1}
        \and P.\,Szkody
        \inst{2}\thanks{Based on observations with the
		Apache Point Observatory (APO) 3.5\,m telescope, which is owned and operated by the
		Astrophysical Research Consortium (ARC)}
        \and J.\,Greiner
        \inst{3}
        \and S.V.\,Zharikov
        \inst{1}
        \and F.-J.\,Zickgraf
        \inst{4}
        \and  A.\,Serrano
        \inst{5}
        \and J.\,Krautter
        \inst{6}
        \and  I.\,Thiering
        \inst{6}
        \and V.\,Neustroev
        \inst{7}\thanks{Special Astron. Obs., 357147 Nizhnij Arkyz, Russia}
          }

   \offprints{G.\,Tovmassian, \\ P.O.Box 439027, San Diego, CA 92143, USA}

   \institute{OAN, Instituto de Astronom\'{\i}a,UNAM, M\'{e}xico\\
              email: gag@astrosen.unam.mx
        \and
              Department of Astronomy, Box 351580,
 University of Washington, Seattle, USA\\
              email: szkody@astro.washington.edu
        \and
             Astrophysical Institute Potsdam, An der Sternwarte 16,
             14482  Potsdam, Germany\\
             email: jgreiner@aip.de
        \and
             Observatoire Astronomique de Strasbourg, France
        \and
             Instituto Nacional de Astrof\'{\i}sica
             Optica y Electr\'onica, AP 51 y 216, Puebla, Pue., M\'{e}xico
        \and
             Landessternwarte
             K\"onigstuhl, 69117 Heidelberg, Germany
        \and
             Udmurtia State University,
             Universitetskaya st.,  Izhevsk, Russia}
%

   \date{Received ; accepted }
        \titlerunning{The New AM Her   System RX J0704.2+6203}
   \maketitle

   \begin{abstract}

We report here on the identification and study of the
 optical counterpart of the  ROSAT source \rxj.
Extensive spectral and photometric observation showed that the object belongs
 to the class
of magnetic Cataclysmic Variables. We determined the orbital period of the
system to be 97\fm27
and estimated the strength of its magnetic field to be on the order of 20 MG. The system was observed
in both high and low states, common for its class. Other parameters of the
 magnetic close binary
system were estimated. The spectral and photometric behavior of
the object is  similar to that of the well studied polar BL Hyi.

   \end{abstract}

%

\section{Introduction}

As part of a program devoted to the optical identification of a complete
sample of northern ROSAT all-sky survey (RASS) X-ray sources, we
identified a new magnetic cataclysmic variable (CV).  Since October 1991
nearly 800 X-ray sources have been observed
within the identification program\footnote{ The identification project is
a collaboration of the Max-Planck-Institut f\"ur extraterrestrische
Physik, Garching, Germany, the Landessternwarte Heidelberg
(LSW), Germany, and the Instituto Nacional de Astrofisica, Optica y
Electronica (INAOE), Puebla, Tonantzintla, Mexico.}.
 A detailed
description of the project is given by Zickgraf \etal\ (\cite{zick97}).
Here we report the identification
and detailed follow-up observations of the optical counterpart of the
ROSAT all-sky survey X-ray source \rxj\, (=  1\,RX\,J070409.2+620330).
It  is the fourth magnetic cataclysmic Variable (CV) discovered within this collaboration
(Tovmassian \etal\ \cite{tov98,tov99,tov00}).


Magnetic CVs (Polars)
are close interacting double systems comprised of a white dwarf (WD)
and a pre-main sequence red dwarf, in which the accretion of matter
onto the WD is governed by its strong magnetic field.
Instead of forming an accretion disc, as common in most other CVs,
matter is funneled to the magnetic pole on the WD through the
magnetic lines.
A standoff shock forms near the surface of the WD. The shocked
plasma cools through bremstrahlung and cyclotron radiation as it
settles onto the WD (Warner \cite{war95}). Polars tend to cluster below the
``period-gap'', i.e. periods less than 2 hours,  and they often exhibit characteristic light curves,
spectral distribution  and profiles of emission lines easily
distinguishable from other types of CVs. BL Hyi is one of the
well studied examples of a polar. We found a number of similarities
in behavior and the parameters of \rxj\, with those of the well studied BL Hyi.
Here we report our observations and  findings on this newly discovered
magnetic CV.


\section {Observations}
\subsection{X-ray observations}

\begin{figure}[t]
\includegraphics[width=6.5cm,angle=-90]{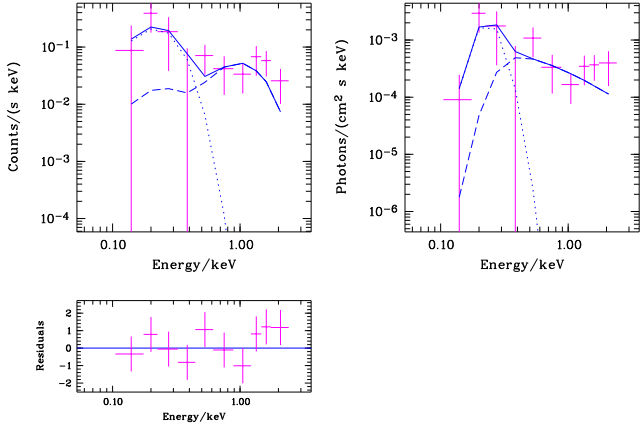}
\caption[]{ROSAT PSPC X-ray spectrum of RX J1554.2+2721 as derived from the all-sky survey
data, fitted with a sum of a blackbody and a thermal 
bremsstrahlung model. The lower left panel shows the deviation between
data and model in units of $\chi^2$ per bin.
}
\label{z1_xspec}
      \end{figure}

RX J0704.2+6203 $\equiv$ 1RXS J070409.2+620330 was scanned during the ROSAT
all-sky-survey over a period of 2.5 days
in Sep. 20-23, 1990 for a total observing time of 340 sec.
Its mean count rate in the ROSAT
position-sensitive proportional counter (PSPC) was 0.13 cts/s,
and the hardness ratio $H\!R1=-0.00\pm0.15$ where $H\!R1$ is defined as 
(H--S)/(H+S), with H (S) being the counts above (below) 0.4 keV over
the full PSPC range of 0.1--2.4 keV.
Thus, the X-ray spectrum is comparatively hard. Despite the small number
of counts a fit with a one-component model, like a pure blackbody model,
is not acceptable. Applying
a sum of a black body and a thermal bremsstrahlung model with the temperature
of the latter component fixed to 20 keV (it is not constrained at all by the 
ROSAT data) gives a good reduced $\chi^2 = 1.04$ (see Fig. \ref{z1_xspec}) 
and the following fit parameters: $kT_{\rm bbdy}$ = 19$\pm$15 eV,
$N_{\rm H}$ = 3.1$\times$10$^{21}$ cm$^{-2}$. For a better comparison to the 
parameters of similar sources we also fixed the black body temperature
to $kT_{\rm bbdy}$ = 25 eV, and derive $N_{\rm H}$ = 
4.5$\times$10$^{20}$ cm$^{-2}$,
Norm$_{\rm bbdy}$ = 0.23 and Norm$_{\rm thbr}$ = 3.0$\times$10$^{-4}$.
This gives an unabsorbed 0.1--2.4 keV flux of 
1.1$\times$10$^{-11}$ erg cm$^{-2}$ s$^{-1}$ 
(or 2.9$\times$10$^{-11}$ erg cm$^{-2}$ s$^{-1}$ bolometric),
corresponding to an unabsorbed bolometric luminosity of 
3.5$\times$10$^{31}$ (D/100 pc)$^2$ erg s$^{-1}$.

\subsection{Optical observations}

     \begin{figure}[t]
\includegraphics[width=9cm,bb=75 455 515 750,clip]{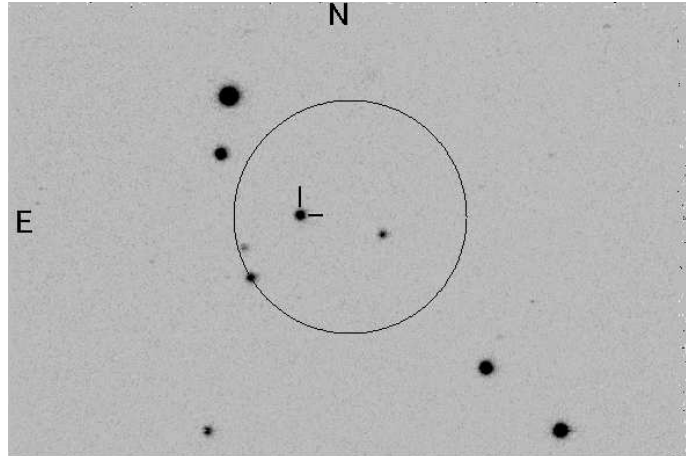}
\caption[]{The CCD  images of the \rxj\ in R$_c$ band  from SAO RAS 1\m telescope.
The object in a high luminosity state at the maximum of the light curve  is marked.
The corresponding ROSAT error box is also marked.
}
\label{fc}
      \end{figure}

The optical counterpart was identified on the 2.1m telescope with the
 faint object spectrograph
(LFOSC, Zickgraf \etal\ \cite{zick97})
at Guillermo Haro Observatory, Cananea, Mexico (Appenzeller \etal\ \cite{app98}).
The finding chart of \rxj\, is presented in Fig.\ref{fc} with the
ROSAT 30\asec  error box marked.
The object was caught in both high (Fig.\ref{fc})  and low  states, common to Polars.
The coordinates of the object derived from POSS are\\
RA= 07$^{\rm h}$ 04$^{\rm m}$ 09\farcs9  DEC=+62\degs 03\amin 27\asec\  Eq.=2000.

\rxj\ was  confirmed as a CV with a few  higher-resolution spectra obtained at
the 2.1m telescope of
Observatorio Astron\'omico Nacional de San Pedro M\'artir (OAN SPM) in Mexico.
The Boller \& Chivens spectrograph with moderate $4-5$\,\AA\  FWHM resolution was
 used. But the  faintness
of the object prevented us from obtaining a  set of data with sufficient time resolution.
Using  the same  telescope with the f/13.5 secondary focus of the 2.1\,m telescope (instead of f/7.5 as used
for spectroscopy),
infrared J  band photometry of the object, covering
multiple orbital periods was accomplished
on Nov 24, 1998. The Camila IR camera operating NICMOS3 $256\times256$ detector sensitive from 1 to 2.5 microns
was used.  IR standard objects were observed  for magnitude calibration.

\rxj\ was observed extensively at the
 1m telescope Zeiss-1000 of the Special Astrophysical
Observatory of Russian Academy of Sciences (SAO RAS)  with the
CCD in Johnson-Kron-Cousins photometric BVRcIc system. In total, the
 observations  span a long
baseline  of several years. Here, again,  we observed standard fields in order to
not only perform differential photometry, but also calculate the
 visual magnitudes of the object.
The log of these and all other optical observations is presented in Table \ref{obstab}.

Observations with low spectral resolution were obtained using the
 long slit spectrograph (UAGS)
at the  prime focus of the  6\,m telescope of SAO RAS on  Nov 04 \& 05  1999.
The data were obtained with $\approx 10$\,\AA\  FWHM resolution in the
 $3630-8340 $\,\AA\ spectral range.
These spectra were flux calibrated using spectrophotometric standard  star
observation during the same
nights and were instrumental for the study of the continuum flux distribution.

Higher spectral resolution observations of \rxj\,  were performed on the
 3.5\,m telescope at
Apache Point Observatory
The double-imaging spectrograph (DIS) was used  in the two wavelength regions 4200-5100 and 5800-6800\,\AA\
to obtain a set of medium-high-spectral resolution (FWHM 2.5\,\AA)  and
high-time resolution (0.1\,P$_{\rm orb}$) spectra  covering almost two orbital periods. Once again
we observed spectrophotometric standards for flux calibration.

\begin{table*}[t]
\caption{The Log of Optical Observations}
\begin{tabular}{llccrrl}
      \noalign{\smallskip}
      \hline
      \noalign{\smallskip}
Date UT & JD & Telescope + Equip. & Filter/Wvlngth & Duration & Exp. & Site \\
     &    &                    &                &  min.     & sec. &     \\
 \noalign{\smallskip}
 \hline
 \noalign{\smallskip}
1997 March 03& 2450510 & 2.1m, B\&Ch+CCD & 3600--6200 & 120 & 1200 & SPM\\
1997 November 01 & 2450754 & 1.0m, CCD & R  & 240 & 600 & SAO\\
1997 November 02 & 2450755 & 1.0m, CCD & R  & 200 & 600 & SAO\\
1998 January 03 & 2450817 & 1.0m, CCD & R  & 140 & 600 & SAO\\
1998 February 27 & 2450872 & 1.0m, CCD & R  & 325 & 600/300 & SAO\\
1998 February 28 & 2450873 & 1.0m, CCD & R  & 185 & 480/300 & SAO\\
1998 March  19& 2450891 & 2.1m, B\&Ch+CCD & 4100--6700 & 120  & 600/900 & SPM\\
1998 November 24 & 2451141 & 2.1m, Camila & J  & 270  & 40 & SPM\\
1998 November 24& 2451141 & 3.5m, DIS & 4240--5060;  & 165 & 600 & APO\\
                &         &            & 5795--6835 &   &     &    \\
1999 January 12 & 2451191 & 1.0m, CCD & V\,I & 285 & 500/240 & SAO\\
1999 January 16 & 2451194 & 1.0m, CCD & R & 210 & 600/300 & SAO\\
1999 January 16 & 2451194 & 1.0m, CCD & B & 25 & 600/300 & SAO\\
1999 November 04 & 2451487 & 6.0m,UAGS+CCD & 3630--8370 & 125 & 600 & SAO\\
1999 November 05 & 2451488 & 6.0m,UAGS+CCD & 3630--8370 & 180 & 600 & SAO\\
       \noalign{\smallskip}
      \hline
    \end{tabular}
   \label{obstab}
   \end{table*}

Both spectral and photometric observations were reduced with  IRAF\footnote {IRAF is the Image Reduction and Analysis Facility, a general purpose
software
             system for the reduction and analysis of astronomical data. IRAF is written and supported by the IRAF
             programming group at the National Optical Astronomy Observatories (NOAO) in Tucson, Arizona. NOAO is
             operated by the Association of Universities for Research in Astronomy (AURA), Inc. under cooperative
             agreement with the National Science Foundation}
packages.

\section {The Orbital Period}
\subsection {Photometric}

The orbital period  of the system was determined based on the full set of
observations that we collected over  3 years. The light curves
in the different bands and in the different luminosity states were all normalized
to the same minimum level. These  also
include fluxes
derived from the spectral observations. The light curves all display distinct
hump structure with varying amplitude depending on wavelength and luminosity
state.  This feature served for period determination
almost as well as an eclipse
does in high--inclination systems. Thus, we were able to derive the orbital
period from photometry with high precision.

We used discrete Fourier transformation techniques for our period search, combined with
the CLEAN procedure (Roberts \etal\ \cite{rob87}), which basically
deconvolves the window
function with  the actual power spectrum in order to depress spectral peaks
originating from the temporal distribution of the data. We first calculated the power spectra of sufficiently
long individual data sets. They all were consistent with each other in showing
the maximum
peak in the power spectral distribution at around 0\fd0675.
Finally we performed the analysis
on the combined set of data.

\begin{figure}
\includegraphics[width=8.8cm]{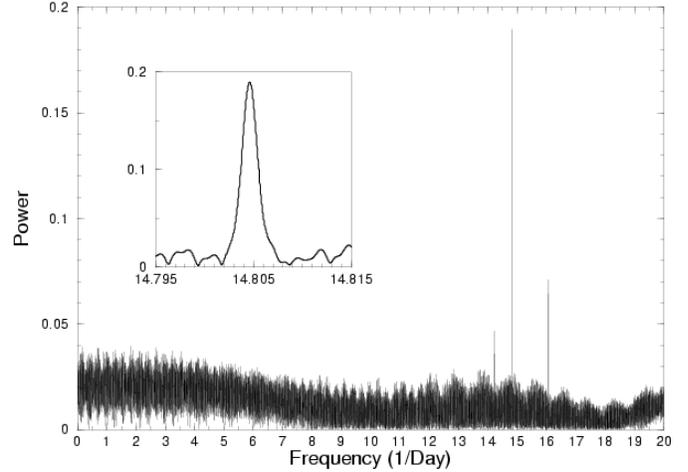}
\caption[]{The CLEANed power spectrum. In the inset the highest
peak of the power spectrum is zoomed up.
}
\label{powsp}
      \end{figure}

Figure \ref{powsp}  presents the CLEANed power
spectrum of the combined data. The strongest peak corresponds to the value
obtained previously from  the separate  sets of data.
The strength  and narrowness of the single peak
(see the inset of the figure for the zoomed up profile of the central peak)
shows the coherence of the periodic signal in the data comprised of measurements
at different wavelengths and even at different luminosity states of the object.
Our estimate is accurate to 0.007
minutes. According to our analysis the orbital period of \rxj\, is
 Porb=0.06754658+/-0.000005 days = 97.2671 min.

\subsection {Spectroscopic}

After  we had a photometric period, we could determine the
spectroscopic period of the system for comparison. We used the set of higher resolution
spectra obtained at the APO 3.5\,m telescope for that purpose.

We used the two strongest lines in the spectrum of \rxj\  (H$\beta$ and \ion{He}{ii}).
Analysis of the profiles of these emission lines has been shown to
be an efficient way to distinguish the various emission components that are present
in magnetic CVs. A brief examination of the
line profiles and constrained two--dimensional images (trailed spectra)
showed that there is a dominant component of the emission lines in both
the Balmer and ionized Helium lines. Thus, it was relatively simple to disentangle
it from the underlying fainter component(s) by simple deblending the
line with two Gaussians. As soon as we had line centers determined for the dominant
component of the emission lines it was easy to fit them with a regular $sin$ curve with
the period determined from photometry. Then the re-calculated values of line
centers of that component from the fit were used for a second iteration of de-blending,
with one Gaussian fixed at the calculated wavelength, to refine   the parameters
of the second component. Although some polars show more than two components
in the emission
lines, in this case, we reached a satisfactory fit and could not find any
evidence of additional sources.    Usually two prominent components are easily
distinguishable. They  are   narrow emission line (NEL) component originating from irradiated  secondary,
and    high velocity component (HVC)  which  probably emitted by  the stream of matter free-falling from
the inner L$_1$ point  toward the WD. The former has FWHM about 2.5\,\AA\ or less, it may reach
up to 400 km/sec semi-amplitude of radial velocity in high-inclination, eclipsing systems, and
its flux is significantly higher at around 0.5 orbital phase, when it  is behind WD and faces observer
with heated side of its surface. While the latter, due to the gradient of intrinsic velocities inside the stream
is broader  and fuzzy (usually about 7-8\,\AA) reaching up to 1500 km/sec in higher inclination systems, but can be much lower
in others. HVC  reaches maximum positive velocity at around orbital phase zero  (see Schwope \etal\ (\cite{sch99}) and references therein
for description of components of emission lines).  Emitting
area of HVC  is less compact and corresponding component in lines of different excitation levels may
have differing phase and semi-amplitude of radial velocities.

\begin{figure}
\includegraphics[width=8.5cm]{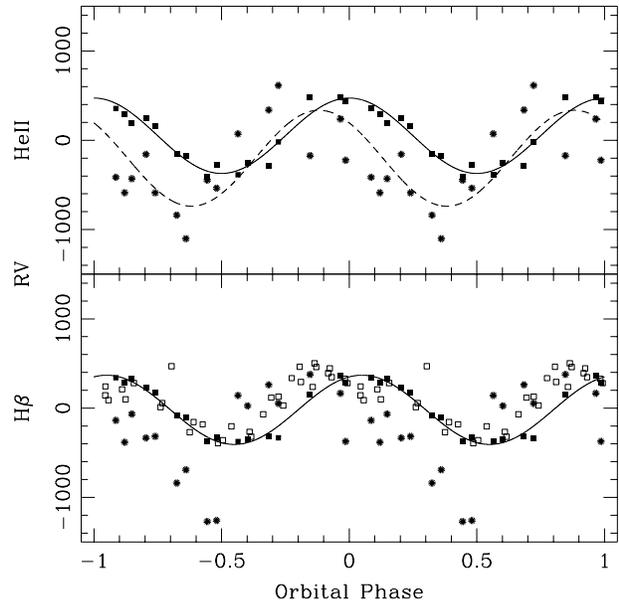}
\caption[]{The radial velocity curves. On the lower panel measurements
of H$\beta$ emission line components  are presented. The stronger and better defined
component marked by filled squares is described by solid $sin$ curve. The second component
supposedly  from the magnetic part of the stream is marked by asterisks. The open squares are
measurements of the whole line on low resolution spectra.

On the upper panel the measurements of both components of \ion{He}{ii} line
are presented with corresponding fits.

}
\label{rv}
      \end{figure}

Thus, the RV of emission lines  were successfully separated into two components
shown with different symbols  in  Figure \ref{rv}.  Both are modulated with
the photometric orbital period. Both show  a relatively large amplitude
of radial velocity variation.  The component drawn by filled squares is
very distinct and is well described by
$$V_{em}=\gamma +
  K_{em1}  \sin(2\pi(t-t_0)/P_{\rm orb})$$

We obtained   K$_{\mathrm {em1}}$= 387 and 422 km/sec for H$\beta$ and
\ion{He}{ii} respectively.
The phase shift of 0.06 P$_{\mathrm {orb}}$ between same component of two different lines is
remarkable in the sense that, the
high values of K$_{\mathrm {em1}}$ combined with the phase shift between lines of various
excitation in a relatively low--inclination system (absence of eclipses),
indicates  that this component of the emission line
probably originates in the stream.
Our designation of this component as pertaining to the stream (HVC) in spite of its relatively low amplitude is
further supported by Doppler tomography (see below).

     \begin{figure*}[ht]
\includegraphics[width=9.2cm]{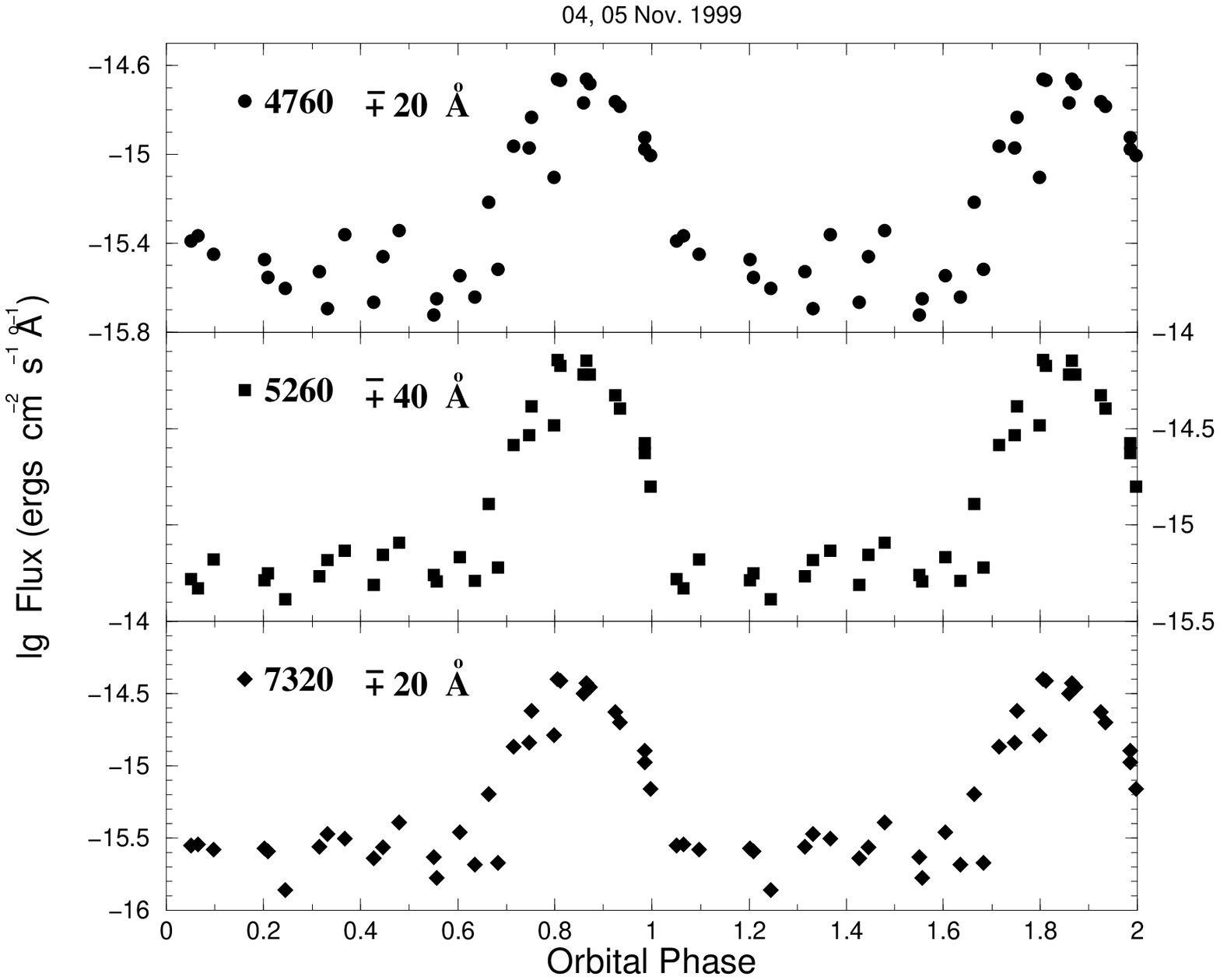}
\includegraphics[width=9cm]{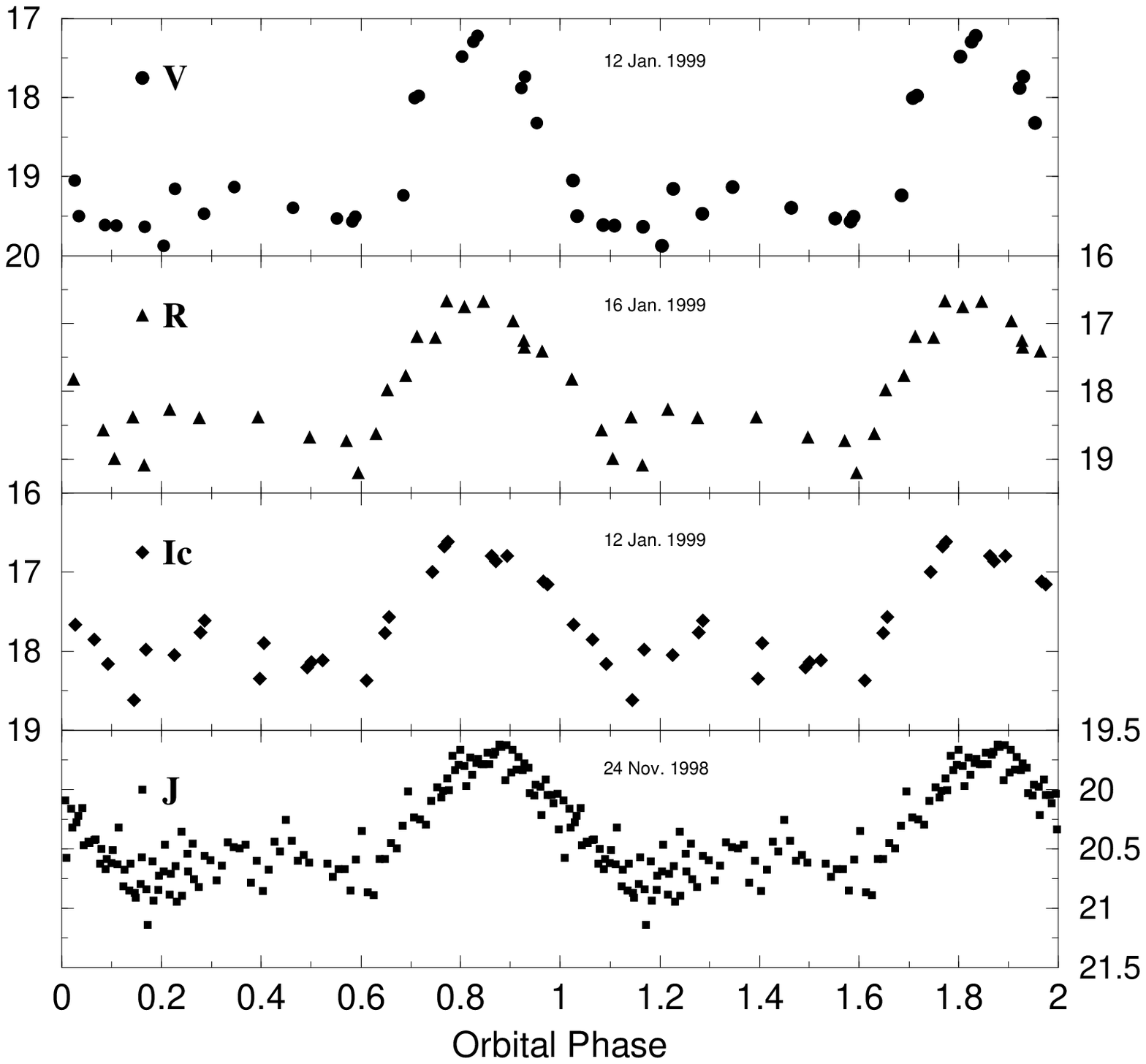}
\caption[]{The light curves of \rxj\  in high luminosity
state obtained at various epochs in different bands (as marked in the panels). 
The left  panel
displays those measured from spectrphotometry in narrow strips of continual
spectrum. The fluxes in log scale are marked on vertical axes. The broad
band photometry presented on right panel are in magnitudes.
}
\label{lc}
      \end{figure*}

The other curve, marked by a dashed  line fitted to the second component of
the emission lines,
is not as well defined. This component fitted  with a sin curve
with semi-amplitude of 540 km/sec supposedly also comes  from the stream,
but from the magnetically  controlled part.
Open squares are measurements of the unblended emission  line in the  low
resolution spectra. They bear signs of both components with predominant
influence of the stronger and  lower velocity  component. We present these measurements
in order to show how good the observations, that are separated by a large time gap,
fold together with the orbital period determined from the photometry.
Note that radial velocity variations are caused by the matter orbiting in the binary system frame,
while humps in the light curve arise from the magnetic pole spinning with WD. Therefore,   the coincidence
of the photometric and spectroscopic periods  provides  good  evidence of synchronized rotation
of the WD in the system i.e.  a polar.

     \begin{figure}[b]
\includegraphics[width=9cm]{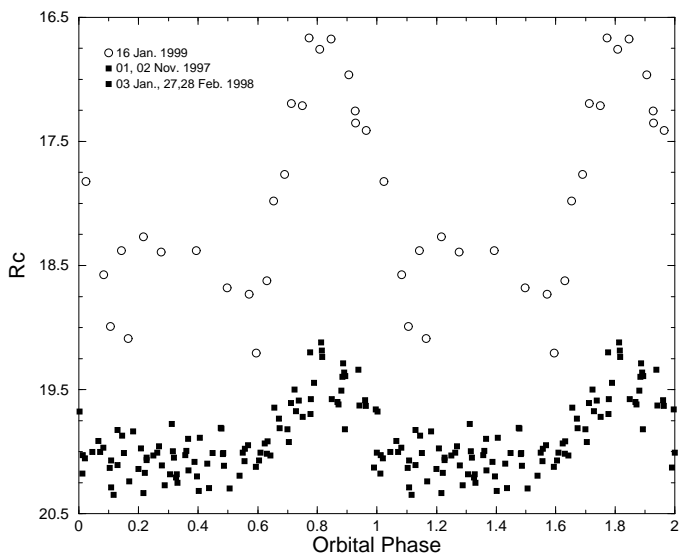}
\caption[]{The light curve of \rxj\ in low luminosity state (filled symbols).
For comparison high state light curve is repeated by open circles. 
Corresponding epochs are indicted in the figure.}
\label{lclow}
      \end{figure}

\section {The nature of the object and its behavior}
\subsection {Photometric characteristics}

The light curves of \rxj\, folded with the orbital period  in different bands
are presented in Figure \ref{lc}. They include light curves derived from spectrophotometric
observations by measuring flux in narrow bands (upper panels) as well as
broad-band CCD photometry (lower).

The light curves exhibit prominent humps that vary insignificantly in amplitude from
band to band and are almost identical in shape.
Although the large number of sub-classes of CVs display a large range
of light curves due to the numerous  sources of radiation, there are easily
distinguishable shapes depending on particular affiliation to one or another class.
The humps in the light curve of \rxj\,  are typical of those  due to a
 large contribution of cyclotron
radiation being beamed from the accretion column. The shape and duration of the humps
suggests that the spot is being eclipsed by the WD itself in its synchronized rotation
with the binary period.
The hump duration is about 0.6 P$_{\mathrm orb}$. There is some indication
 of a secondary  peak at phase 1.4,
which may be attributed to the presence of the second weak accretion pole in the system.
The light curve is remarkably similar to that of BL Hyi (Wolff \etal\ \cite{wol99}).

\rxj\, was observed in R$_{\mathrm c}$ band in a high luminosity state as well
as in a low state.
The phase folded light curve in the low state is presented in Figure \ref{lclow}
along with the high state light
curve in the same filter for  comparison. Interestingly, the
light curve shape in the low state remains remarkably similar  to the shape in high state.
However, the amplitude of the cyclotron humps  in the  low state is  a factor
of two  less
than the in high state.

\subsection {Spectrophotometric characteristics}

\begin{figure}
\includegraphics[width=9cm]{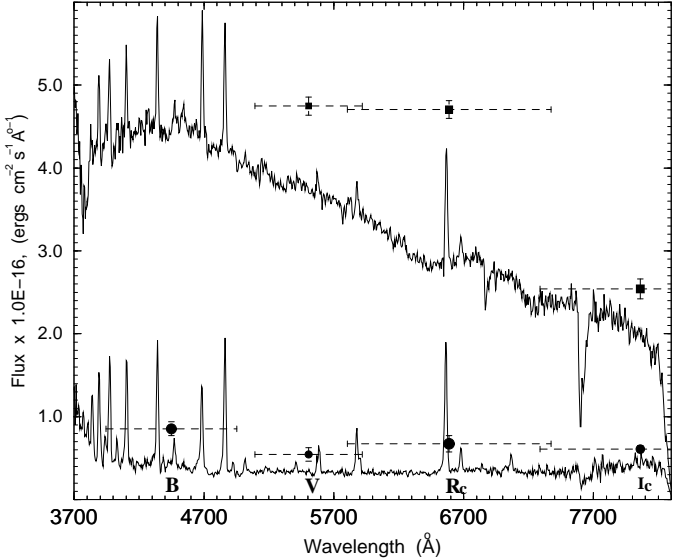}

\caption[]{Spectra of \rxj\ at the top of the cyclotron hump and at the bottom.
}
\label{splow}
      \end{figure}

The low resolution spectra of \rxj\,  obtained at the 6m telescope SAO RAS were
used
for a study of the continuum variation over the orbital period. As can be seen from flux measurements
in narrow bands covering the continuum spectrum in Fig.\ref{lc}, the spectrum
undergoes large
changes in the course of the  orbit. It can be better demonstrated by comparison of the
averaged spectrum of \rxj\, at the maximum of the light  curve humps with the
 averaged spectrum
at the bottom (see Fig. \ref{splow}. There is an enormous change in shape of the spectra,
confirmed by the multiwavelength photometry also shown in the Figure.

Besides the usual set
of Balmer and He lines in emission common for CVs, the spectrum at the maximum
 of the light curve
shows a steep increase  in the blue and small humps  in the red.
No traces of
lines from the secondary could be found.
Change of shape of the spectrum  and continuum  humps  are  common for
magnetic CVs. Similar spectral changes were observed in another magnetic CV
recently reported by Tovmassian \etal\  (\cite{tov00}).
Particularly, the  spectra of \rxj\,  very much resembles
those of the  well studied BL Hyi (Schwope \etal\  \cite{sch95}).

The humps seen in the continuum of  magnetic CVs are actually lines of
cyclotron emission emitted from the magnetic column of matter accreting onto the magnetic pole of
the WD in the binary system.
The cyclotron emission  originates
in the column of  in-falling
matter above the magnetic pole, where it  slows down and heats  up before settling on the
surface of the WD.
 It is emitted
 in a wide range of wavelengths from the near UV to the infrared region.
Since cyclotron radiation is beamed perpendicular to the magnetic lines and
since the WD is locked by its magnetic field in a  synchronous rotation with
the binary system, we see periodic maxima   in the light curve and cyclotron
spectrum appear and disappear depending on the viewing angle of  the magnetic
pole of the WD.

\begin{figure}[t]
\includegraphics[width=9cm]{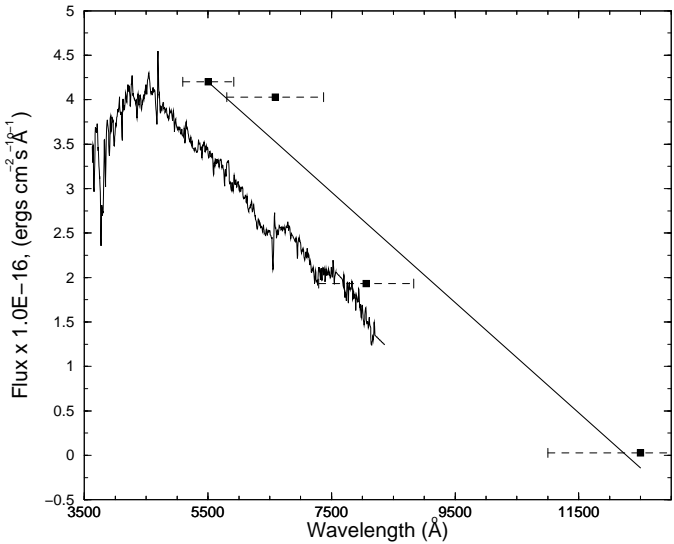}

\caption[]{Differential spectrum of \rxj\ between the maximum and minimum
in the light curve from spectrophotometric and photometric data (straight line).
Cyclotron humps are clearly seen in the red part of the spectrum
}
\label{difsp}
      \end{figure}

In order to separate the bulk of  cyclotron emission from the rest of the emitting sources
 and estimate the strength of the
magnetic field, the averaged
spectrum at the minimum was subtracted from the spectrum at the  maximum of the
 light curve.
The spectra were shifted according  to radial velocity measurements corresponding
to the orbital motion.
The residual spectrum is presented  in  Figure \ref{difsp}. The lines are clearly gone,
thus their contribution to the luminosity change is negligible. The
continuum bears apparent signs of cyclotron lines seen as humps.

We used the theory  developed by Chanmugam \& Dulk (\cite{cd81}), Wickramasinghe \& Meggitt  (\cite{wime85})
to  constrain our models in  order to describe the  observed differential  flux of  \rxj\,
assuming its cyclotron nature. Similar models were applied by Schwope \etal\  (\cite{sch95})
and Ferrario \etal\  (\cite{fer96}) to estimate the magnetic field of the WD in BL Hyi.
We calculated  a set of models for   emission cyclotron spectra that rises from
the  slab of  plasma representing  the post-shock  region  of the accretion
column to fit our observations. We took the simplest
view assuming a homogeneously emitting  plasma.

Assuming a typical value of the order of the shock temperature for $kT = 20$\,keV
the observed  cyclotron lines in the red portion of the spectrum
can be identified as 8,7 \& 6 harmonics (corresponding to peaks at $\approx$ 7450, 6600 \& 5950 \AA)
of  cyclotron emission, thus leading to a moderate 20\,MG magnetic field strength.
Since cyclotron emission is described by a
number of parameters (temperature $kT$, the polar angle between the line of sight
and the magnetic field $\theta$, the field strength $B$ and a plasma
parameter $\Lambda = l n_e/B$, where $l$ is  geometric size, $n_e$ -- electron density,
(Wickramasinghe \& Meggitt \cite{wime85}), there is some ambiguity of chosen parameters and
they can be traded against each other (e.g., lower
values of $B$ and $T$ require higher values of $\Lambda$ in order to
shift the cyclotron peak to the observed value and vice verse).
Alternatively, the  observed lines  could be 9,8 \& 7 harmonics (a shift by one),
which will require lower value of $kT=10$\,keV in order to successfully fit the
cyclotron spectrum, which in turn will lead to an increment of the
 magnetic field  up to 24\,MG.
The difference is not significant, taking into account the much wider range of magnetic fields
observed in mCVs.
The suggested rather low value of the magnetic field strength is
supported by the relatively hard X-ray spectrum typically found in
low-field polars (Beuermann \& Burwitz \cite{bebu};  Schwope \cite{schw96}).

Meanwhile, the thick straight line  in the Fig.\ref{difsp} is a linear fit to the  differential energy distribution
between the maximum  and minimum  of the light curve, derived from photometry. It extends beyond
the optical range thanks to our measurements in near IR-band\,J.
Its steepness is an additional confirmation of the  cyclotron nature of the
 observed radiation.

Thus, our tentative classification of \rxj\, as a member of AM\,Her class of magnetic
Cataclysmic Variables based on the shape of the lightcurve, presence of significant X-ray radiation and
 strong \ion{He}{ii} line, characteristic profiles of emission lines and the  semiamplitude of RV,
as well as defined high and low luminosity states, is confirmed by direct estimate of the magnetic field of
WD by fitting  the cyclotron emission spectrum. We can go one step further and estimate other parameters of the system.
>From the fact that the light curves show no eclipses, other  than self eclipse of the
accreting, magnetic pole by the WD itself, we can safely assume that the
inclination of the system is less than
$\approx70\degs$. On the other hand we can measure substantial high velocities from the stream of transferred
matter (upto 420 km/sec). At least the ballistic part of the stream should lie in or close to the orbital
plane, thus the inclination of the system can not be very small, otherwise projected velocities would be
much less than those observed in high inclination systems like HU  Aquarii (Schwope \etal\  \cite{sch97})
and RX\,J0719.2+6557 (Tovmassian \etal\ \cite{tov99}).
So we anticipate that the inclination of the system is not less than  $30\degs$. This in turn will lead to
restrictions imposed on an angle $\beta$  between the rotation axis and line of sight. The fact that
the magnetic pole remains behind the limb of the WD during almost 0.6  orbital phase, as evident  from
the light curves,  the angle $\beta$ should be somewhere in between 120 and 150 degrees, as follows from the
definition of the magnetic  geometry by Cropper(\cite{cro90}).

\section {Doppler tomography}

    \begin{figure*}[t]
\includegraphics[width=18cm]{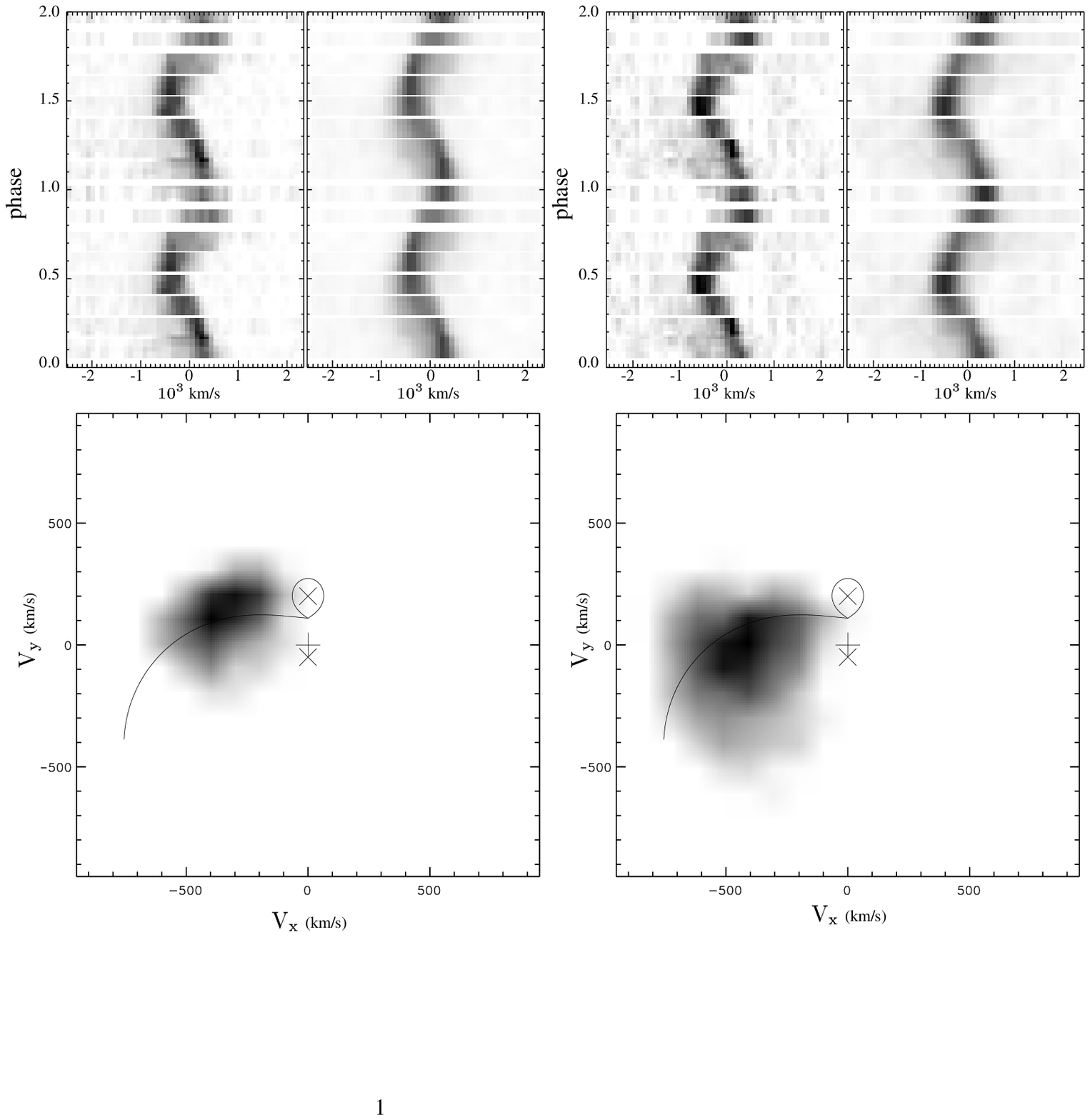}

\caption[]{The Doppler maps  of  the  H$\beta$ (left) and \ion{He}{ii} (right) emission lines of \rxj.
The secondary star Roche lobe and ballistic accretion stream trajectory are plotted for an assumed
 inclination angle of 30\degs, primary mass M$_1$=0.7 and mass ratio   $q=0.25$.
 Above panels show phase folded  observed  and reconstructed   trailed   spectra of corresponding lines.
}
\label{dopmap}
      \end{figure*}

Doppler tomography (Marsh \& Horne \cite{maho88}) is a technique that allows
a determination of
the velocity field of the matter in the orbital plane of close binary systems.
It has been  successfully applied in a number of studies of CVs and particularly
magnetic systems (i.e. Schwope \etal\ \cite{sch97}).
Usually by providing orbital phases as a parameter for  Doppler reconstruction,
one obtains velocity maps or the images in a fixed coordinate system. Our examination of light
curves and radial velocity curves failed to determine the
orbital phases of the binary system. the absence of eclipses or any measurable contribution
from the stellar components of the system prevents us from using  common methods
of phase determination.
Nevertheless we performed Doppler tomography on our data using an
 arbitrary  zero
point for orbital phases. We used an optimized maximum entropy method (MEM; the code kindly provided by
Spruit \cite{spruit}). The resulting images clearly show the prolonged concentration of emission (matter),
that is usually attributed to the stream from the donor secondary star to the
WD (see for example Hoard \etal\ \cite{hoa99}).
We estimated the zero point of orbital phases that placed the accretion stream
at the usual location in the Doppler map. The corresponding  inferior conjunction is at $T0 = 2451141.87 \pm 0.005$.
Although the uncertainty of such
a determination, based on eye inspection, is high (up to 0.1 orbital  phase),
its validity is out of doubt.
Therefore, MEM Doppler maps confirm that the emission lines originate in the accretion
stream, with the free-fall component being a major contributor. The resulting maps
are presented in Figure \ref{dopmap} with corresponding  observed and reconstructed 
trailed spectra of
emission lines.  Note the difference in velocities of \ion{He}{ii} and H$_\beta$!\,  
There is more emission coming from magnetic part of the stream in \ion{He}{ii} than
in H$_\beta$. In H$_\beta$ tomogram we actually see the main component of the emission line,
coming from the balistic part, because it is much stronger than the second component.
While in \ion{He}{ii} the second component has more contribution and easily could be traced
by eye  in the phase-folded trailed spectrum (between phases 1.0 -- 1.5) as well as on the 
velocity map. It is seen as extention of the main spot towards negative V$_y$ velocities. 
For indication of
Roche lobe of secondary,  location of primary WD and the stream of matter   in the map
we selected reasonable  binary parameters (the inclination angle of 30\degs, primary mass M$_1$=0.7 with mass ratio $q=0.25$)  to fit the observed characteristics
of the stream.

 The absence of evidence for  secondary heating in \rxj\  in the  Doppler maps
is noteworthy.
The single  spot on the maps clearly can not  belong to the irradiated secondary
 due to its  extremely high velocity
and  its  elongated shape. This particular feature is another characteristic that makes \rxj\ a
twin of BL Hyi (Mennickent \etal \cite{men}), since the majority of AM Her
systems show a significant contribution from the
irradiated secondary in the line emission.

\section {Conclusions}

We found that the optical counterpart of X-Ray source
\rxj\  is  a magnetic CV (AM\,Her type).

The orbital period of the system is 97 min and it increments
the number of polars that are  clustered around a 100 min peak.

The preliminary estimate of magnetic strength of probably dipole
WD in the system is B=20-24\,MG.

The system resembles the southern
polar BL Hyi  in its characteristics. The most striking similarity is that both
 systems have no
signs of an irradiated secondary star in their emission lines,  unlike
other AM Her objects. (Mennickent etal \cite{men})

our rough estimate of the inclination angle is  $30<i<60$
and of the angle between the  rotation axis and field vector is
 $120< \beta <150$.

Photometry obtained in low and high states shows that the magnetically
driven accretion does not halt but reduces its magnitude at  the  low state.
The reduction of cyclotron
luminosity by $\approx2$ magnitudes leads to the conclusion that either accretion
area or accretion  rate  decreases by a factor of 6.

The absence of secondary irradiation does not support speculations that
the accretion rate changes could be the result of the heating of a part of
that star.

\begin{acknowledgements}

 GT is supported by CONACYT under  grant 25454-A. PS acknowledges partial support from
NASA LTSA grant NAG-53345. JG is supported by the Deutsche Agentur f\"ur
Raumfahrtangelegenheiten (DARA) GmbH under contract FKZ 50 QQ 9602 3.
The ROSAT  project is supported by the German Bundes\-mini\-ste\-rium f\"ur
Bildung, Wissenschaft, Forschung und Technologie (BMBF/DARA) and the
Max-Planck-Society.

\end{acknowledgements}

\end{document}